\documentclass[aps,prd,10pt,twocolumn,superscriptaddress,showpacs,footinbib]{revtex4-1}
\pdfoutput=1
\usepackage{hyperref}
\usepackage{slashed}
\usepackage{graphicx}
\usepackage{enumerate}
\usepackage{amsfonts,amsmath,amssymb,bm,bbm}
\usepackage{color}
\usepackage{epsfig, subfigure}
\usepackage{setspace}
\usepackage{booktabs, tabularx} 
\usepackage{multirow}

\hypersetup{
    pdfnewwindow=true,      % links in new window
    colorlinks=true,       % false: boxed links; true: colored links
    linkcolor=blue,          % color of internal links
    citecolor=blue,        % color of links to bibliography
    filecolor=blue,      % color of file links
    urlcolor=blue        % color of external links
}

\newcommand{\be}{\begin{equation}}  
\newcommand{\ee}{\end{equation}}
\newcommand{\bea}{\begin{eqnarray}}  
\newcommand{\eea}{\end{eqnarray}}

\newcommand{\nue}{\nu_e}

\newcommand{\nux}{\nu_x}

\newcommand{\nueb}{\bar{\nu}_e}

\begin{document}

\title{Gadolinium in Water Cherenkov Detectors Improves Detection of Supernova $\nu_e$}

\author{Ranjan Laha}
\affiliation{Center for Cosmology and AstroParticle Physics (CCAPP), Ohio State University, Columbus, OH 43210}
\affiliation{Department of Physics, Ohio State University, Columbus, OH 43210}

\author{John F. Beacom}
\affiliation{Center for Cosmology and AstroParticle Physics (CCAPP), Ohio State University, Columbus, OH 43210}
\affiliation{Department of Physics, Ohio State University, Columbus, OH 43210}
\affiliation{Department of Astronomy, Ohio State University, Columbus, OH 43210\\
{\tt laha.1@osu.edu,beacom.7@osu.edu} \smallskip}

\date{\today}

%%%%%%%%%%%%%%%%%%%%%%%%%%%%%%%%%%%%%%%%%%
\begin{abstract}
Detecting supernova $\nu_e$ is essential for testing supernova and neutrino physics, but the yields are small and the backgrounds from other channels large, e.g., $\sim 10^2$ and $\sim 10^4$ events, respectively, in Super-Kamiokande. We develop a new way to isolate supernova $\nu_e$, using gadolinium-loaded water Cherenkov detectors.  The forward-peaked nature of $\nu_e + e^- \rightarrow \nu_e + e^-$ allows an angular cut that contains the majority of events.  Even in a narrow cone, near-isotropic inverse beta events, $\bar{\nu}_e + p \rightarrow e^+ + n$, are a large background.  With neutron detection by radiative capture on gadolinium, these background events can be individually identified with high efficiency.  The remaining backgrounds are smaller and can be measured separately, so they can be statistically subtracted.  Super-Kamiokande with gadolinium could measure the total and average energy of supernova $\nu_e$ with $\sim$ 20\% precision or better each (90\% C.L.).  Hyper-Kamiokande with gadolinium could improve this by a factor of $\sim$ 5.  This precision will allow powerful tests of supernova neutrino emission, neutrino mixing, and exotic physics.  Unless very large liquid argon or liquid scintillator detectors are built, this is the only way to guarantee precise measurements of supernova $\nu_e$.  
\end{abstract}
%%%%%%%%%%%%%%%%%%%%%%%%%%%%%%%%%%%%%%%%%%

\maketitle

%%%%%%%%%%%%%%%%%%%%%%%%%%%%%%%%%%%%%%%%%%%%%%%%%%%
\section{Introduction}
\label{sec:introduction}
%%%%%%%%%%%%%%%%%%%%%%%%%%%%%%%%%%%%%%%%%%%%%%%%%%%

Supernovae are one of the most spectacular electromagnetic displays in the Universe.  Understanding them is essential for many areas of physics and astrophysics.  Core-collapse supernovae are massive stars ($\gtrsim 8M_{\odot}$) that, at the end of their burning cycles, collapse under gravity to form a neutron star or black hole~\cite{Janka:1995bx,Langanke:2002ab,Mezzacappa:2005ju,Burrows:2006ci,Janka:2006fh,Woosley:2006ie,Burrows:2012ew,Janka:2012wk}.  These collapses are potential sites for gravitational-wave production~\cite{Ott:2008wt,Kotake:2011yv,Ando:2012hna}, gamma-ray bursts~\cite{Woosley:2006fn}, heavy-element nucleosynthesis~\cite{Thielemann:2001rn,Woosley:2007as}, and cosmic-ray acceleration~\cite{Blasi:2010gr}.

It is difficult to learn about the core properties and collapse mechanism using electromagnetic light curves, as the surface of last scattering of photons is in the outer envelope.  Neutrinos, on the other hand, being weakly interacting, have their surface of last scattering much deeper inside, within the core.  Neutrinos carry about $\sim 99\%$ of the binding energy released during the collapse of the star.  Precise measurements of all flavors of neutrinos can provide much information about a supernova~\cite{Thompson:2002mw,Tomas:2004gr,Dasgupta:2005wn,Friedland:2006ta,Lunardini:2007vn,Suwa:2008sf,Marek:2008qi,Huedepohl:2009wh,Kneller:2010sc,Muller:2011yi,Kistler:2012as,Lund:2012vm,Wongwathanarat:2012zp,Scholberg:2012id,Tamborra:2013laa,Nakamura:2012nc,Kneller:2013baa,Kneller:2013ska,Borriello:2013tha}.

The only supernova neutrinos ever detected were from SN 1987A~\cite{Hirata:1987hu,Bionta:1987qt}.  Even this modest data has been invaluable for understanding neutrinos and supernovae.  Only $\nueb$ were detected, through the inverse beta channel, $\bar{\nu}_e + p \rightarrow e^+ + n$, leading to, e.g., constraints on the total and average energy in this flavor~\cite{Jegerlehner:1996kx,Lunardini:2004bj,Yuksel:2007mn,Pagliaroli:2008ur}.  (We assume that the first event was not due to neutrino-electron elastic scattering, which has a very small probability.)

Computer simulations of supernova explosions have detailed predictions about the neutrino emission, but, due to the lack of a high-statistics Galactic supernova, it is not possible to adequately test these~\cite{Mueller:2010nf,Brandt:2010xa,Mueller:2012ak,O'Connor:2012am,Kotake:2012iv,Ott:2012kr,Kotake:2012nd,Mueller:2012is,Bruenn:2012mj}.  It is important to detect all flavors of neutrinos to measure the total and average energy in each.  Because the differences between flavors may be modest, large numbers of events must be detected to ensure adequate precision.

Galactic supernovae occur only once every $\sim 30$ years.  It is essential that a variety of detectors be ready to detect all flavors of neutrinos well to understand the physics and astrophysics of core-collapse supernovae.  Using present detectors, it will be easy to measure supernova $\bar{\nu}_e$ and $\nu_x$, via inverse beta and elastic scattering on protons, respectively~\cite{Vogel:1999zy,Strumia:2003zx,Beacom:2002hs,Dasgupta:2011wg}. Unless very large liquid argon~\cite{Raghavan:1986hv,GilBotella:2004bv} or liquid scintillator detectors~\cite{Wurm:2011zn,Li:2013zyd} are built, or other techniques become experimentally viable~\cite{Haxton:1987bf,Fuller:1998kb,Haxton:1998vj,Ianni:2005ki,Lazauskas:2009yh,Suzuki:2012aa}, there is presently no way to guarantee the clean detection of supernova $\nu_e$ in adequate numbers.  The difficulties of measuring the $\nu_e$ spectrum well enough have long been known; e.g., see Refs.~\cite{Haxton:1987kc, Minakata:1989mk,Qian:1993hh,Lunardini:2001pb,Lunardini:2003eh,Beacom:2005it,Skadhauge:2006su,Skadhauge:2008gf}.

Here we show how this problem could be solved by using gadolinium (Gd) in Super-Kamiokande (Super-K) and other large water Cherenkov detectors.  The addition of Gd to Super-K was proposed to improve the detection of $\bar{\nu}_e$.  Ironically, this would also improve the detection of $\nu_e$.  We add new ideas to those briefly noted in Ref.~\cite{Beacom:2003nk} and perform the first detailed calculations, showing how supernova $\nu_e$ could be measured precisely.

The principal technique is to use neutrino-electron scattering, $\nu_e + e^- \rightarrow \nu_e + e^-$.  These events are forward-peaked, so a narrow cone contains the majority of them.  The largest background is from inverse beta events.  The use of Gd to detect neutrons will help in individually detecting and removing these events with high efficiency.  The spectrum of $\bar{\nu}_e$ will be measured precisely so that the remaining inverse beta and $\bar{\nu}_e + e^-$ scattering events can be statistically subtracted from the forward cone.  Liquid scintillator detectors can detect $\nux$ (= $\nu_{\mu} + \nu_{\tau}$) well enough through $\nu + p \rightarrow \nu + p$ scattering, so the $\nu_x + e^-$ scattering events can be statistically subtracted.

In addition, we show how gadolinium will improve the prospects for measuring $\nu_e$ charged-current interactions with oxygen.  This channel is only important if the average energy of $\nu_e$ is large, either intrinsically, or due to efficient mixing with sufficiently hot $\nu_x$.  Recent supernova simulations suggest that none of the flavors has a large average energy, and that the differences between flavors are modest, so that these interactions with oxygen may not be important.  In contrast, the neutrino-electron scattering events would be measured well in all scenarios if Gd is used to reduce backgrounds.

Detecting supernova $\nu_e$ will be helpful in constructing the initial spectrum of these neutrinos, testing neutrino mixing scenarios, and probing exotic physics.  We concentrate on detecting the $\nu_e$ emitted during the full duration of the burst; however, this technique could also help in detecting the short neutronization burst $\nu_e$ in Mton water Cherenkov detectors~\cite{Kachelriess:2004ds}.

The outline for this article is as follows.  In Sec.~\ref{sec:Neutrino spectra and Gd}, we discuss the various theoretical and experimental inputs required to isolate supernova $\nue$.  In Sec.~\ref{sec:SN nue detection}, we discuss how this can constrain the $\nu_e$ spectrum parameters, and we conclude in Sec.~\ref{sec:conclusion}.

%%%%%%%%%%%%%%%%%%%%%%%%%%%%%%%%%%%%%%%%%%%%%%%%%%%
\section{Calculation inputs}
\label{sec:Neutrino spectra and Gd}
%%%%%%%%%%%%%%%%%%%%%%%%%%%%%%%%%%%%%%%%%%%%%%%%%%%

We first discuss the neutrino spectra from a supernova, followed by the various detection channels in a water Cherenkov detector.  We then outline the detection strategy that we propose to use to detect supernova $\nu_e$ in a water Cherenkov detector with gadolinium.

\subsection{Supernova Neutrino Spectra}

A supernova neutrino burst lasts for $\sim$ 10 sec and includes all flavors of neutrinos.  The total binding energy released in the explosion is $\sim 3 \times 10^{53}$ erg.  We assume that the total energy is equipartitioned between the 6 species so that the total energy carried by each $\nu$ (or $\bar{\nu}$) flavor is $\sim 5 \times 10^{52}$ erg. The supernova is assumed to be at a distance of 10 kpc, the median distance of core collapse progenitor stars in our Galaxy, which is slightly farther than the distance to the Galactic Center~\cite{Adams:2013ana}.

Supernova neutrinos are emitted in a quasi-thermal distribution.  For concreteness, we take a particular modified Maxwell-Boltzmann spectrum~\cite{Keil:2002in,Tamborra:2012ac}, 
\begin{equation}
f(E_\nu)=\frac{128}{3}\frac{E_\nu ^3}{\langle E_\nu \rangle ^4} \,{\rm exp} \left({-\frac{4E_\nu}{\langle E_\nu \rangle}}\right)\, ,
\label{eq:normalised spectra}
\end{equation} 
where this is normalized to unity.  Using a regular Maxwell-Boltzmann or a Fermi-Dirac spectrum with the same average energy gives more neutrinos at high energies.  For the electron-scattering and inverse-beta channels, the increased number of events is $\lesssim$ 5\%.  For the oxygen channel, which depends very sensitively on neutrino energy, the number of events can increase by $\sim$ 50\%.  Our choice of spectrum is conservative and our results can only improve if other neutrino spectra are appropriate.

For the average energies of the initial spectra, we take $\langle E_{\nu_e} \rangle \approx$ 11 -- 12 MeV, $\langle E_{\bar{\nu}_e} \rangle \approx$ 14 -- 15 MeV, and $\langle E_{\nu_x} \rangle \approx$ 15 -- 18 MeV; the hierarchy follows from the different strengths of interaction in the supernova core.  Neutrino mixing effects in the supernova~\cite{Dighe:1999bi,Dighe:2008dq,Dasgupta:2009mg,Dasgupta:2010gr,Duan:2010bg,Duan:2010af,Friedland:2010sc,Pehlivan:2011hp,Sarikas:2011am,Cherry:2012zw} or in Earth~\cite{Lunardini:2001pb,Dighe:2003jg,Dighe:2003vm} can have a dramatic effect on the final spectra, even exchanging them.  Then the $\nu_e$ (or $\nueb$) spectrum could have an average energy of $\sim$ 15 -- 18 MeV, increasing the yields of charged-current detection channels.  (The yields of neutral-current detection channels do not change for active-flavor mixing.)  To tell how efficient the mixing is, we need to measure the $\nu_e$ detection spectra precisely.

A model independent neutrino signal from a supernova is the neutronization burst, which consists of a short pulse ($\sim$ 25 msec) of initially pure $\nu_e$ before the $\sim$ 10 sec emission of neutrinos of all flavors~\cite{Kachelriess:2004ds}.  Depending on the neutrino mixing scenario, the number of neutronization $\nu_e$ detected in a Mton water Cherenkov detector for a Galactic supernova is $\sim$ 30 -- 100~\cite{Kachelriess:2004ds, Cherry:2013mv}.  Our detection strategy will also be useful in this case.  In Super-Kamiokande (fiducial volume 32 kton), the number of events due to neutronisation $\nu_e$ is only $\sim$ $\mathcal{O}$(1).

\subsection{Neutrino Detection Interactions}

All flavors of neutrinos and antineutrinos can be detected with the $\nu + e^- \rightarrow \nu + e^-$ channel.  The recoil kinetic energy of the scattered electron varies between 0 and $2E_\nu ^2/(m_e+2E_\nu)$.  The forward-scattered electron makes an angle $\alpha$ with the incoming neutrino given by cos $\alpha=\sqrt{T_e/(T_e+2 m_e)}(E_\nu + m_e)/E_\nu$, where $T_e$ is the kinetic energy of the recoil electron.

The differential cross section for neutrino-electron elastic scattering is~\cite{Vogel:1989iv} 
\begin{eqnarray}
 \frac{d\sigma}{dT_e}&=&\frac{G_F^2 m_e}{2\pi}\bigg[(g_V+g_A)^2+(g_V-g_A)^2 \left(1-\frac{T_e}{E_{\nu}}\right)^2 \nonumber\\
 &+&(g_A^2-g_V^2)\frac{m_eT_e}{E_{\nu}^2}\bigg]\, ,
 \end{eqnarray}
 \linebreak
where $G_F$ is the Fermi coupling constant, $g_V=2$ sin$^2\theta_W \pm 1/2$ for $\nu_e$ and $\nu_x$, respectively, and $g_A=\pm 1/2$ for $\nu_e$ and $\nu_x$, respectively.  For anti-neutrinos, $g_A \rightarrow -g_A$.  When integrated over $T_e$, the total cross section $\sigma(E_\nu) \propto m_e E_\nu$.

Only $\nueb$ were detected from SN 1987A, via the inverse beta reaction, $\bar{\nu}_e + p \rightarrow e^+ + n$, where $p$ denotes free hydrogen (protons) in water and the positrons are emitted almost isotropically.  The cross section for this process is $\sigma (E_\nu) \simeq 0.0952 \times 10^{-42}  (E_\nu -1.3)^2 (1 - 7 E_\nu/m_p)$ cm$^2$ where $m_p$ is the proton mass, the energies are in MeV, the threshold of the reaction is $E_\nu > 1.8$ MeV, and $T_e \simeq E_\nu - 1.8$ MeV~\cite{Vogel:1999zy,Strumia:2003zx}.

The neutron thermalizes by elastic collisions and is captured on protons as $n + p \rightarrow d + \gamma$ in about 200 $\mu s$.  The emitted gamma ray has an energy of 2.2 MeV, which cannot be reliably detected in Super-K due to low-energy detector backgrounds~\cite{Zhang:2013tua}.  To unambiguously detect the emitted neutron, it has been proposed to add Gd to large water Cherenkov detectors.  Then the neutron will be thermalized and captured on Gd in about 20 $\mu s$, leading to a 3 -- 4 gamma rays with a total energy of about 8 MeV, which is easily detectable in Super-K~\cite{Beacom:2003nk}.

Electron neutrinos can also be detected in water Cherenkov detectors by $\nu_e + ^{16}$O $\rightarrow e^- + ^{16}$F$^*$~\cite{Haxton:1987kc}, where most of the final-state decay products of the excited $^{16}$F$^*$ nucleus are not detectable.  The threshold for this reaction is $\approx$ 15 MeV, and the electron kinetic energy is $T_e \approx E_\nu -15$ MeV.  In the energy range $25 \, {\rm MeV} \leq E_\nu \leq 100$ MeV, the cross-section is given by $\sigma(E_\nu) \approx 4.7 \times 10^{-40} ( E_\nu^{0.25} -15^{0.25})^6$ cm$^2$, for energies in MeV~\cite{Haxton:1987kc, Tomas:2003xn}.  The angular distribution of the electrons is slightly backward tilted.  The steep energy dependence of the cross section means that $\nu_e$ can only be detected well if the average energy is large, say due to mixing.

We neglect other neutrino interactions with oxygen ($\bar{\nu}_e$ charged-current~\cite{Haxton:1987kc} and all-flavor neutral-current~\cite{Langanke:1995he}), as they are not our focus and their yields are small compared to that from the inverse beta channel.

%%%%%%%%%%%%%%%%%%%%%%%%%%%%%%%%%%%%%%%%%%
\begin{table}[b]
\caption{Expected numbers of events in Super-K for a Galactic supernova at a distance of 10 kpc for different values of the neutrino average energy (we do not round the numbers so that small differences remain visible).  The total energy of the supernova is assumed to be $3 \times 10^{53}$ erg, equipartitioned among all flavors (here $\nu_x = \nu_\mu + \nu_\tau$).  The detection threshold during a burst is assumed to be $T_e = 3$ MeV.  Other interactions with oxygen are neglected because their yields are small compared to that of inverse beta decay.}
\setlength{\extrarowheight}{4pt} 
\begin{ruledtabular}
\begin{spacing}{1.1}
\begin{tabular}{lccc}
Detection channel & 12 MeV & 15 MeV & 18 MeV \\ 
\hline
$\nu_e + e^- \rightarrow \nu_e + e^-$ & 188 & 203 & 212 \\
$\bar{\nu}_e + e^- \rightarrow \bar{\nu}_e + e^-$ & 56 & 64 & 70 \\
$\nu_x + e^- \rightarrow \nu_x + e^-$ & 60 & 64 & 68 \\
$\bar{\nu}_x + e^- \rightarrow \bar{\nu}_x + e^-$ & 48 & 54 & 56 \\
\hline
$\nu_e + ^{16}$O $\rightarrow e^- + ^{16}$F$^*$ & 16 & 70 & 202 \\
\hline
$\bar{\nu}_e + p \rightarrow e^+ + n$ & 5662 & 7071 & 8345 \\
\end{tabular}
\end{spacing}
\end{ruledtabular}
\label{tab:yields}
\end{table}
%%%%%%%%%%%%%%%%%%%%%%%%%%%%%%%%%%%%%%%%%%

The time-integrated flux for single neutrino flavor is
 \begin{equation}
\frac{dF}{dE_{\nu}}=\frac{1}{4\pi d^2}\frac{E_{\nu}^{\rm tot}}{\langle E_{\nu} \rangle}f(E_{\nu})\, ,\end{equation}
where $E_{\nu}^{\rm tot}$ denotes the total energy in that $\nu$ flavor and $d$ is the distance to the supernova.  The observed event spectrum in the detector is
\begin{equation}
\frac{dN}{dT_e}=N_T \, \int _{E_{\rm{min}}} ^{\infty} dE_{\nu} \, \frac{dF}{dE_{\nu}} (E_\nu) \, \frac{d\sigma}{dT_e}(E_{\nu}, T_e)\, ,
\end{equation}
where $N_T$ is the appropriate number of targets.  For a larger average energy, the thermally-averaged cross section is larger, but the flux is smaller (because the total energy is taken to be fixed).  For neutrino-electron scattering, these effects nearly cancel, making the total number of events almost insensitive to the average energy.  The shape of the electron recoil spectrum does change, which provides sensitivity to the average energy.

Table~\ref{tab:yields} shows the expected number of events in Super-K for these reactions under different assumptions about the neutrino average energy.  For additional details about the detection of neutrinos from a Galactic supernova in water Cherenkov detectors, see the references already cited as well as Refs.~\cite{Beacom:1998ya,Beacom:1998yb,Ikeda:2007sa}.

\subsection{Proposed Detection Strategy}

%%%%%%%%%%%%%%%%%%%%%%%%%%%%%%%%%%%%%%%%%%%%%%%%%%%
\begin{figure}[t]
\centering
\includegraphics[angle=0.0,width=0.48\textwidth]{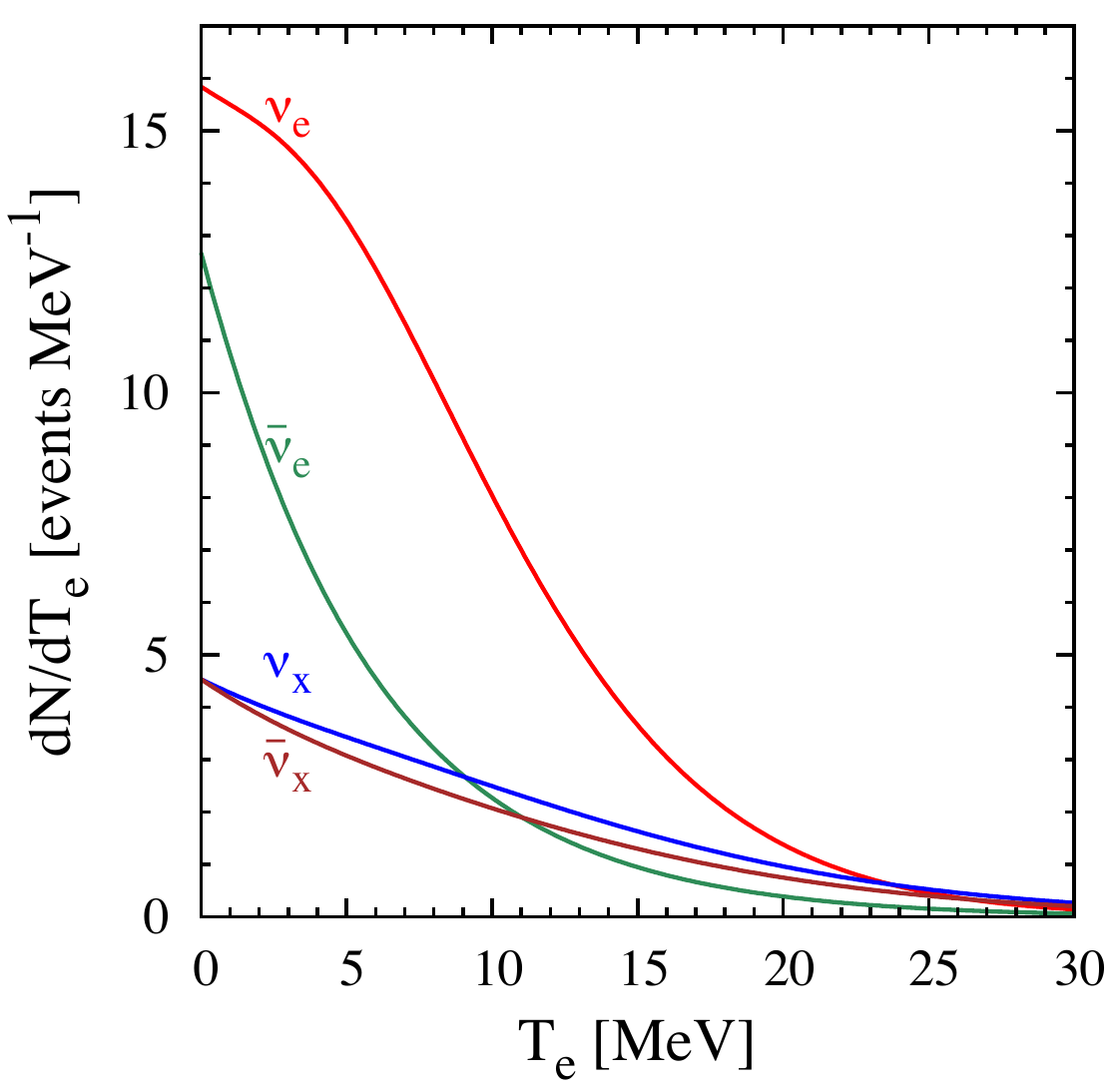}
\caption{Electron spectra for the $\nu + e^- \rightarrow \nu + e^-$ detection channels for a supernova in Super-K.  These are just the events in the forward 40$^\circ$ cone ($\sim 68\%$ of the total).  We take $\langle E_{\nu_e} \rangle = 12$ MeV, $\langle E_{\bar{\nu}_e} \rangle = 15$ MeV, and $\langle E_{\nu_x} \rangle = 18$ MeV; the other assumptions are listed in Table~\ref{tab:yields}.}
\label{fig:nu - e event spectrum}
\end{figure}
%%%%%%%%%%%%%%%%%%%%%%%%%%%%%%%%%%%%%%%%%%%%%%%%%%%

%%%%%%%%%%%%%%%%%%%%%%%%%%%%%%%%%%%%%%%%%%%%%%%%%%%
\begin{figure*}[t]
\centering
\includegraphics[angle=0.0,width=0.48\textwidth]{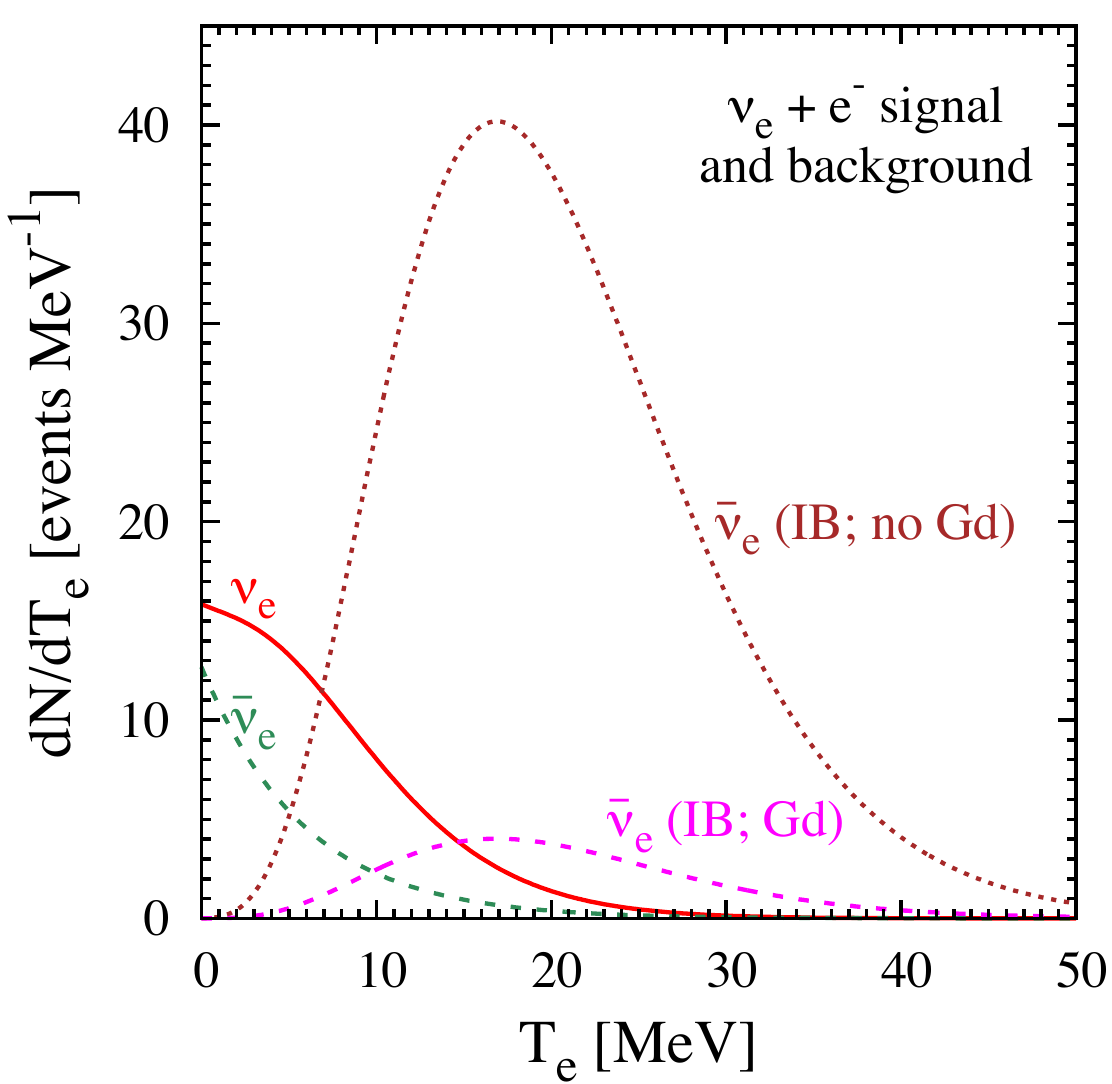}
\includegraphics[angle=0.0,width=0.48\textwidth]{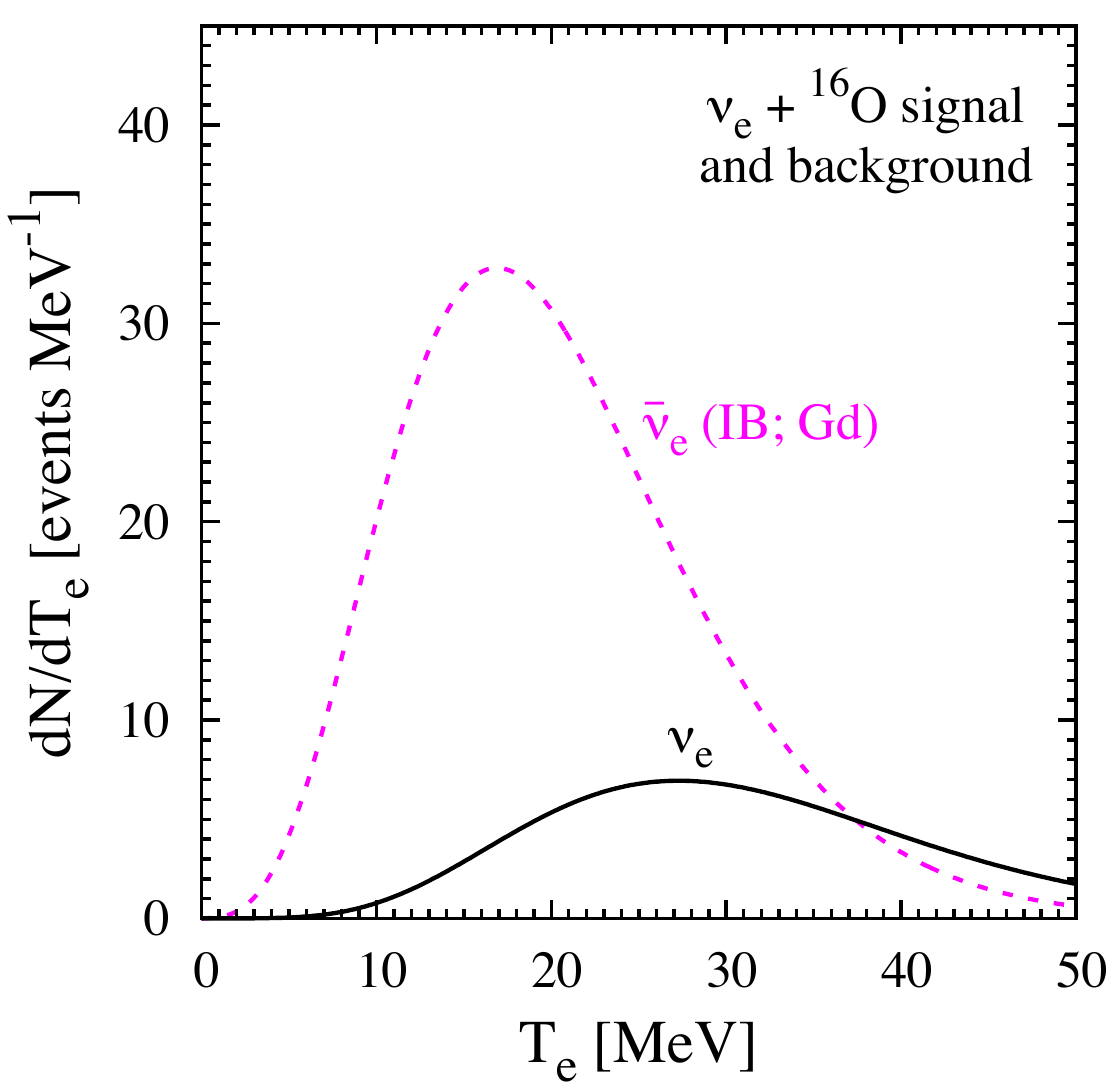}
\caption{Detectable electron (or positron) spectra in Super-K without or with Gd.  The two panels consider different cases for $\langle E_{\nu_e} \rangle$ after neutrino mixing.  Other parameters, including $\langle E_{\bar{\nu}_e} \rangle = 15$ MeV, are as in Fig.~\ref{fig:nu - e event spectrum}.  {\bf Left Panel:} For Case (A) with $\langle E_{\nu_e} \rangle = 12$ MeV, we focus on the $\nu_e + e^-$ signal (solid line) in the forward 40$^\circ$ cone.  The dotted line shows the large inverse beta background without Gd, and the dashed lines show the most important backgrounds with Gd.  {\bf Right panel:}  For Case (B) with $\langle E_{\nu_e} \rangle = 18$ MeV, we focus on the $\nu_e + ^{16}$O signal (solid line) in the region complementary to the forward 25$^\circ$ cone (note the different angle).  The inverse beta background without Gd is too large to show, and dashed line shows this background with Gd.  Here the signal and background are both due to the Galactic supernova.}
\label{fig:IB}
\vspace{0.1cm}
\end{figure*}
%%%%%%%%%%%%%%%%%%%%%%%%%%%%%%%%%%%%%%%%%%%%%%%%%%%

%%%%%%%%%%%%%%%%%%%%%%%%%%%%%%%%%%%%%%%%%%%%%%%%%%%
\begin{figure*}[t]
\centering
\includegraphics[angle=0.0,width=0.48\textwidth]{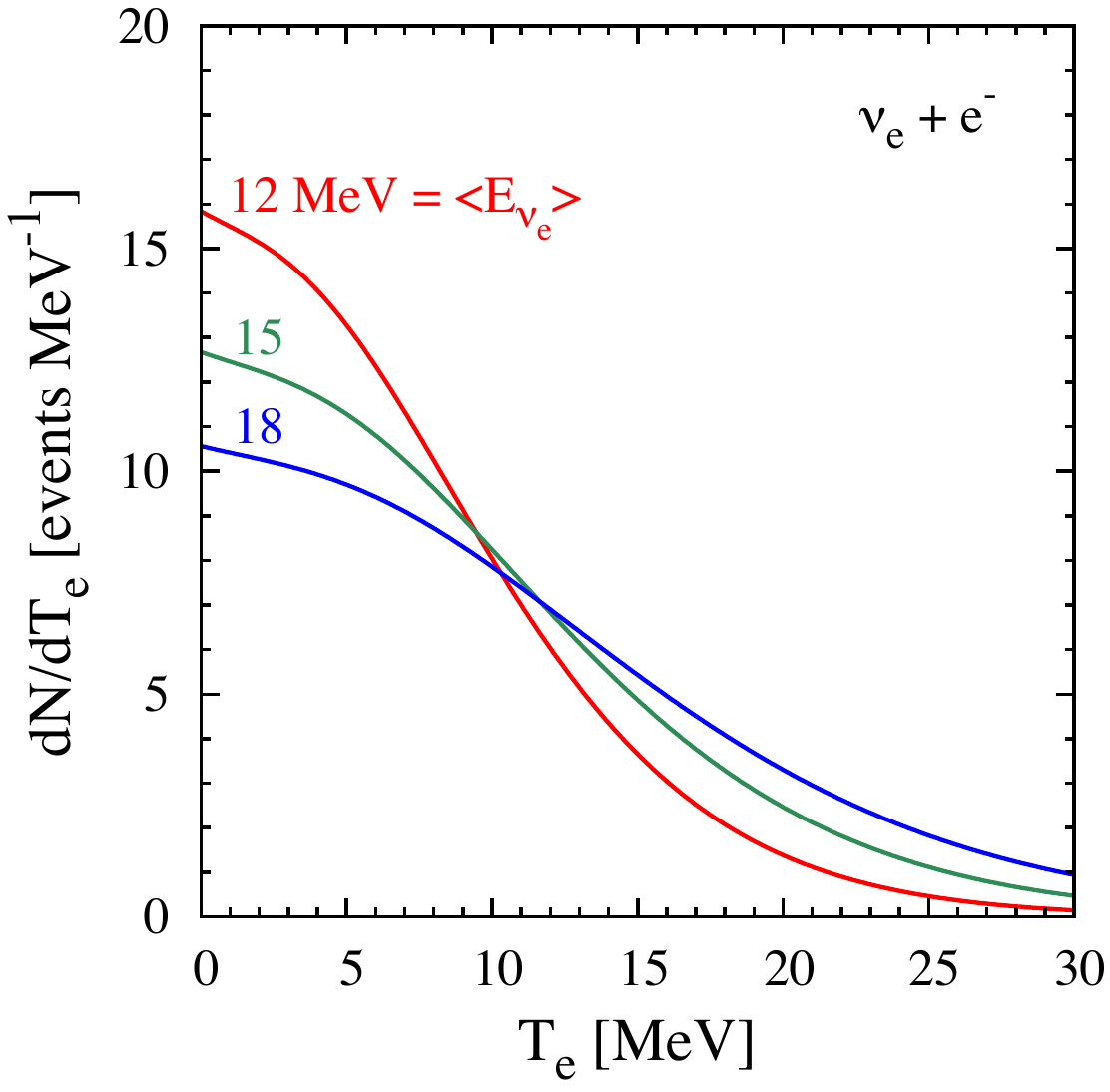}
\includegraphics[angle=0.0,width=0.48\textwidth]{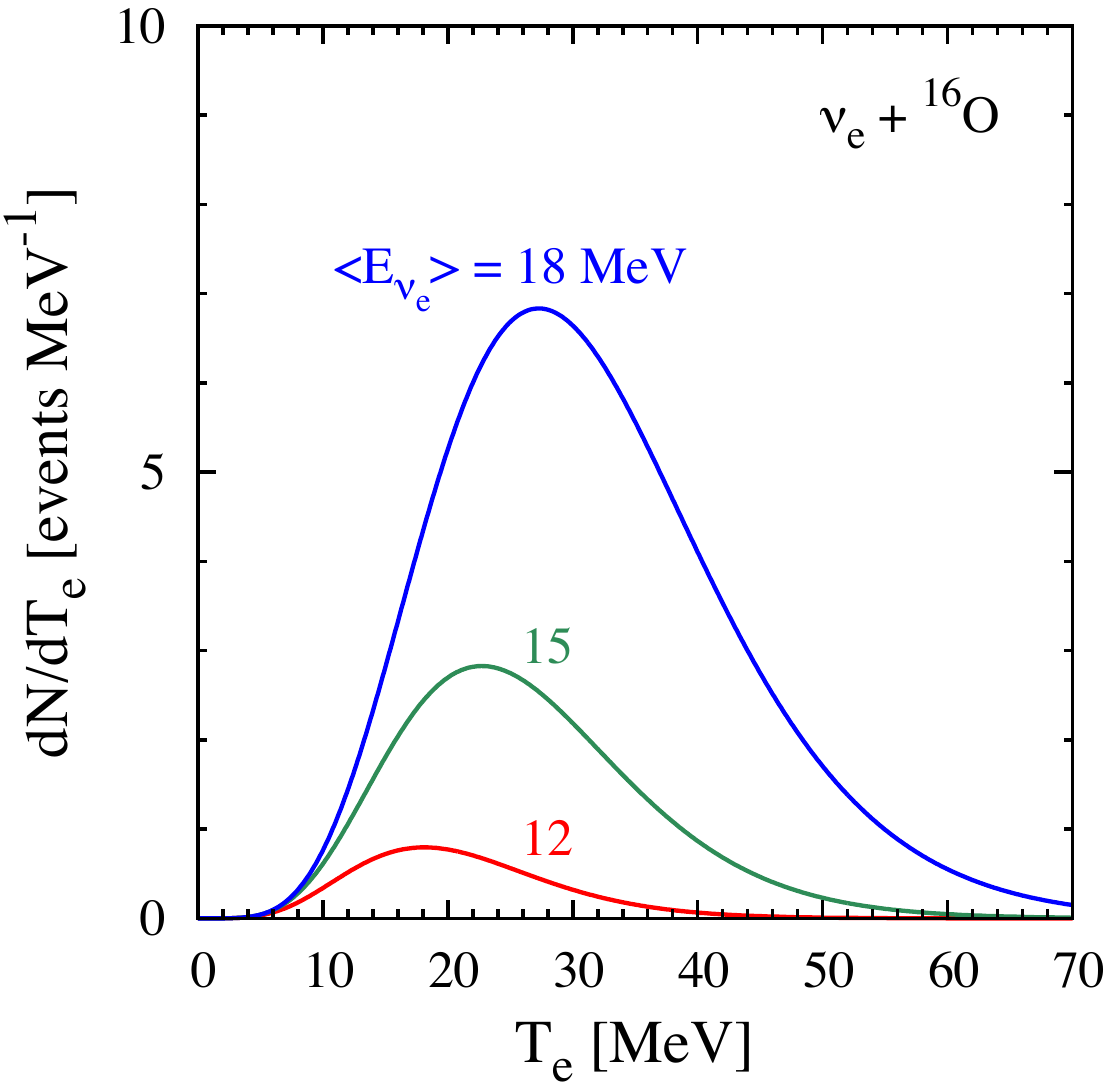}
\caption{Detectable electron spectra in Super-K, ignoring backgrounds, for different assumed average energies for $\nu_e$ (12, 15, and 18 MeV) to show variants of the signals in Fig.~\ref{fig:IB}.  All spectra scale linearly with changes in the assumed total energy in $\nu_e$.  Other assumptions as above.  Note axis changes from Fig.~\ref{fig:IB}.  {\bf Left Panel:} For the $\nu_e + e^-$ channel in the forward 40$^\circ$ cone.  {\bf Right Panel:} For the $\nu_e + ^{16}$O channel in the region complementary to the forward 25$^\circ$ cone.}
\label{fig:different average energies}
\end{figure*}
%%%%%%%%%%%%%%%%%%%%%%%%%%%%%%%%%%%%%%%%%%%%%%%%%%%

%%%%%%%%%%%%%%%%%%%%%%%%%%%%%%%%%%%%%%%%%%%%%%%%%%%
\begin{figure*}[t]
\centering
\includegraphics[angle=0.0,width=0.48\textwidth]{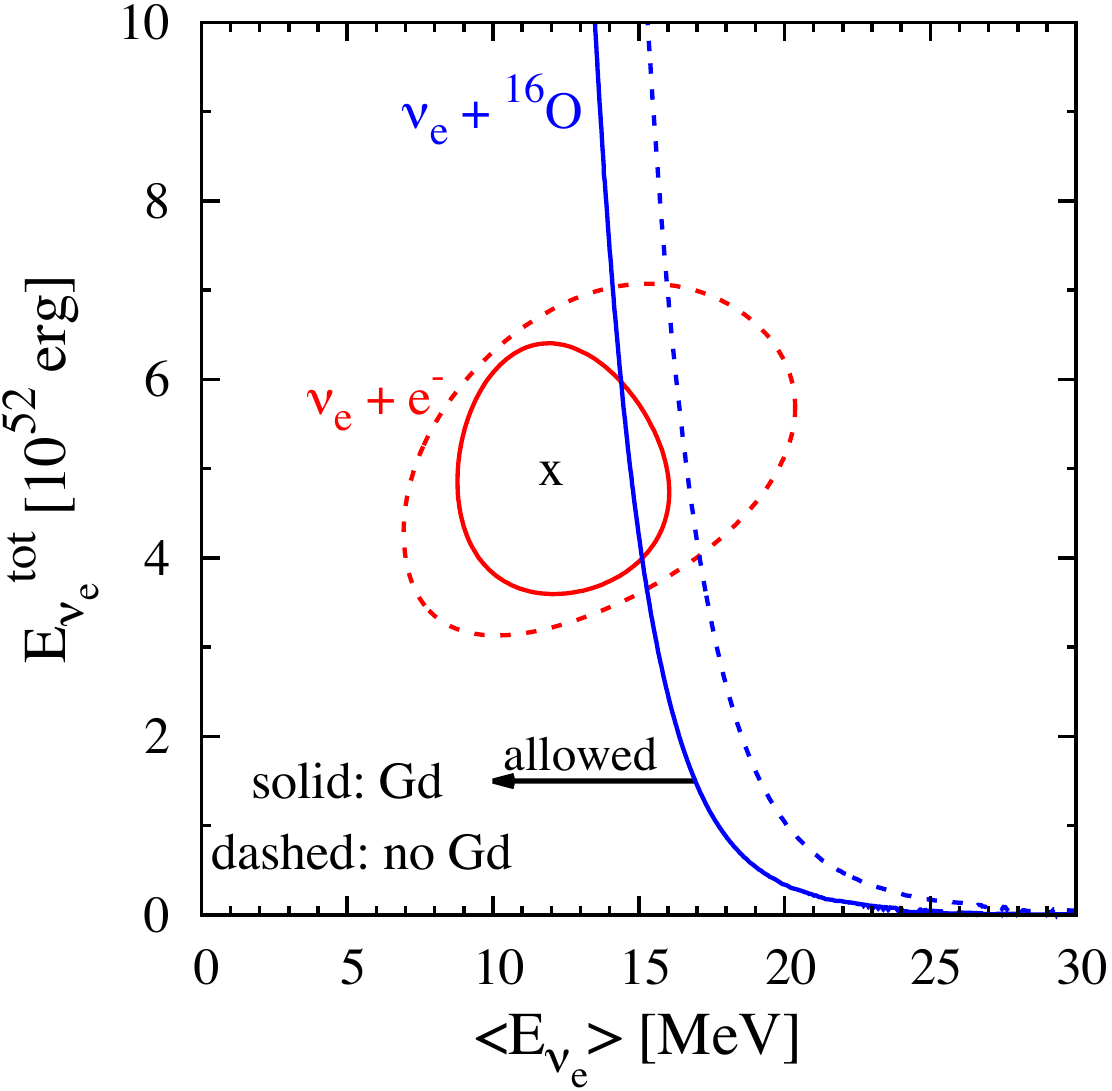}
\includegraphics[angle=0.0,width=0.48\textwidth]{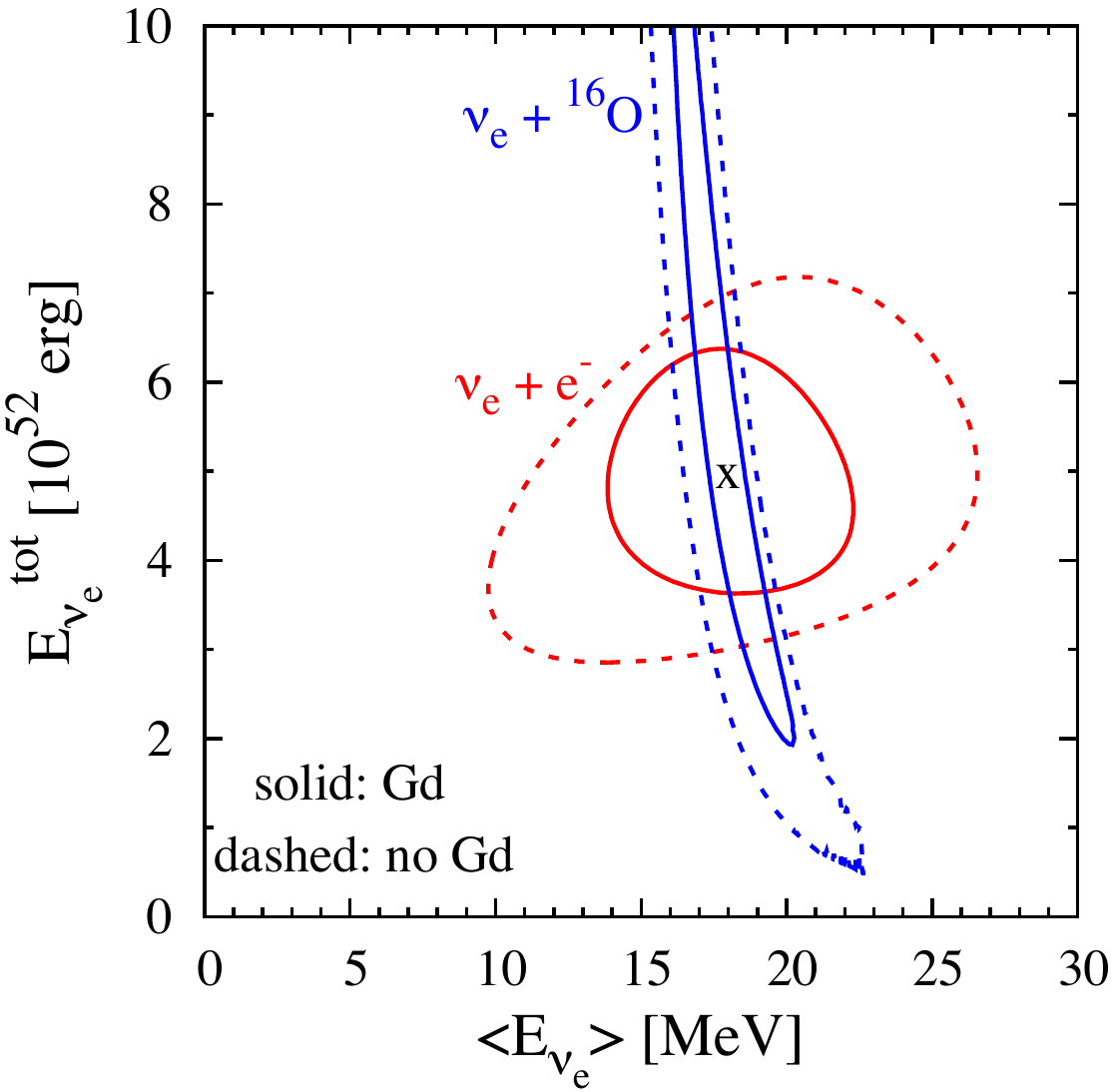}
\caption{Allowed regions (90\% C.L. $\Delta \chi ^2$ contours) for the $\nu_e$ spectrum parameters determined from the $\nu_e + e^-$ and $\nu_e + ^{16}$O channels separately.  The combined constraints (not shown) closely follow what would be expected visually.  The two panels are for different cases (fiducial parameters marked by an x), matching those of Fig.~\ref{fig:IB}.  Dashed lines indicate the contours when Gd is not used, and solid lines show the improvements when Gd is used.  {\bf Left Panel:} When the $\nu_e$ average energy is small, here 12 MeV, the $\nu_e + e^-$ channel gives a closed allowed region but the $\nu_e + ^{16}$O channel only defines upper limits.  {\bf Right Panel:} When the $\nu_e$ average energy is large, here 18 MeV, both channels give closed allowed regions.}
\label{fig:chisquared}
\end{figure*}
%%%%%%%%%%%%%%%%%%%%%%%%%%%%%%%%%%%%%%%%%%%%%%%%%%%

%%%%%%%%%%%%%%%%%%%%%%%%%%%%%%%%%%%%%%%%%%%%%%%%%%%
 \begin{figure}[t]
\centering
\includegraphics[angle=0.0,width=0.48\textwidth]{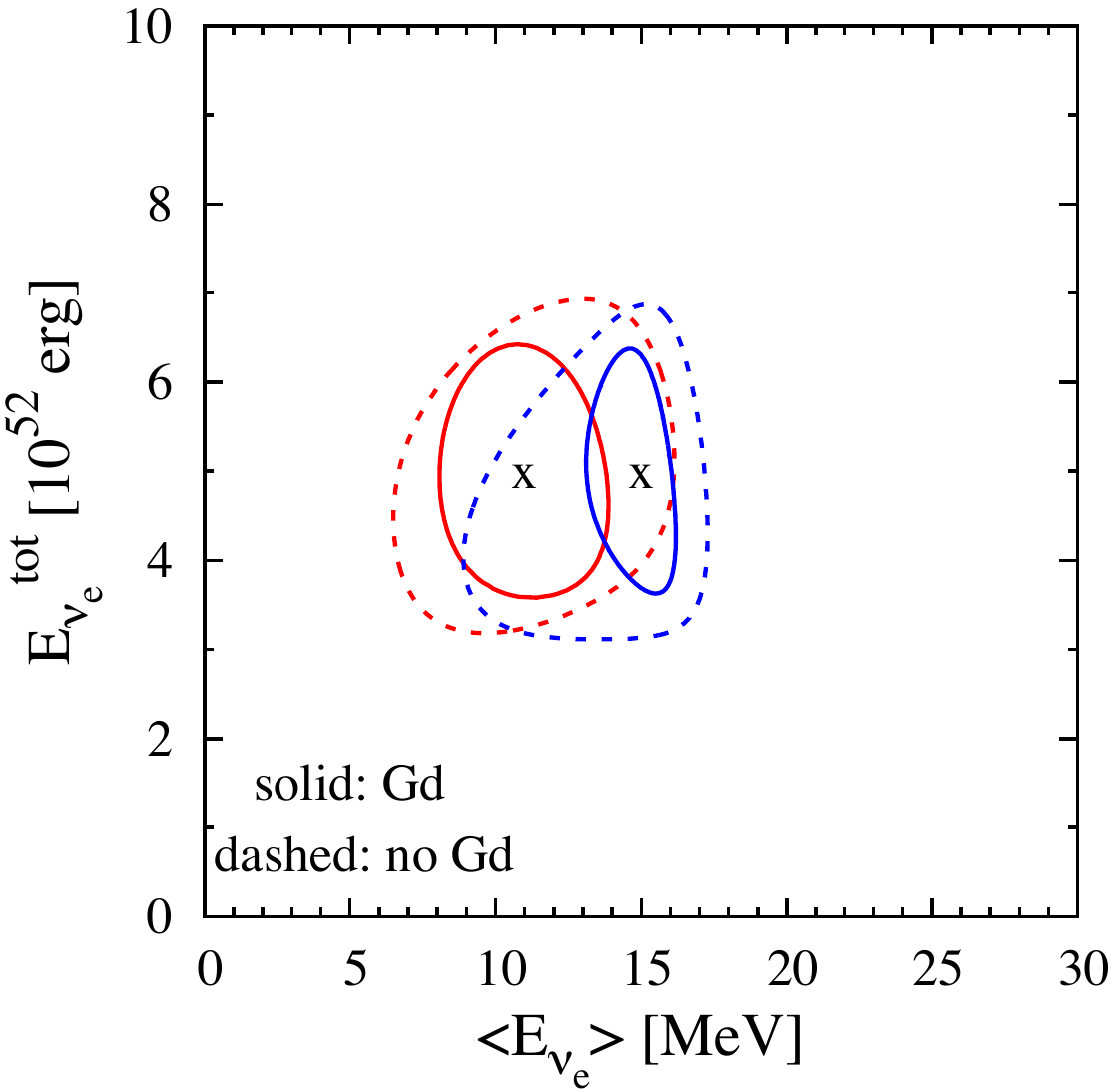}
\caption{Allowed regions (90\% C.L. $\Delta \chi ^2$ contours) for the $\nu_e$ spectrum parameters determined from the $\nu_e + e^-$ and $\nu_e + ^{16}$O channels jointly.  Two examples of fiducial parameters ($\langle E_{\nu_e} \rangle^0 = 11$ MeV and $\langle E_{\nu_e} \rangle^0 = 15$ MeV) are each marked with an x.  The corresponding fit regions are shown without and with Gd.}
\label{fig:jointchisquared}
\end{figure}
%%%%%%%%%%%%%%%%%%%%%%%%%%%%%%%%%%%%%%%%%%%%%%%%%%%

We focus on Super-K, the largest detector with low intrinsic backgrounds~\cite{Abe:2010hy}.  We assume that supernova events can be detected in the full inner volume of 32 kton.  Super-K measures the energy, position, and direction of charged particles with very high efficiency.  During a burst, detector backgrounds can be ignored.  There is extensive ongoing research on employing Gd in Super-K~\cite{Beacom:2003nk,Watanabe:2008ru,Vagins:2012vta}.  The efficiency of neutron capture on Gd will be known from calibration data.

We employ the $\nu_e + e^- \rightarrow \nu_e + e^-$ reaction to detect the $\nu_e$ and look for the forward-scattered electrons.  Knowing the direction of the Galactic supernova, if we make an angular cut of half-angle 40$^{\circ}$ (appropriate for the lowest energy $\nu_e + e^-$ events~\cite{Abe:2010hy}), then $\sim$ 68\% of the electron-scattering events will be in that cone.  The forward-scattered electrons can also locate the supernova to within a few degrees~\cite{Beacom:1998fj,Tomas:2003xn,Antonioli:2004zb,Adams:2013ana}.

Fig.~\ref{fig:nu - e event spectrum} shows the recoil spectra for neutrino-electron scattering for all flavors.  (We use kinetic energy, but Super-Kamiokande conventionally uses the total energy, $E_e = T_e + m_e$).  Because the energy range is so broad, the effects of energy resolution smearing ($\sim 15\%$ near 10 MeV) were found to be modest, and are not included.  As can be seen from the figure, $\nu_e$ has the largest number of events.  This is important, because the other flavors of neutrino-electron scattering events are an irreducible background to the $\nu_e + e^-$ events.

The largest number of events will be due to the inverse beta reaction, which is almost isotropic.  Neutron detection on Gd will individually identify $\sim$ 90$\%$ of these events.  The very large number of events will determine the $\bar{\nu}_e$ parameters precisely ($\sim 1\%$ with $\sim 10^4$ events), which will be used to statistically subtract the remaining inverse beta events.  Events from other detection channels can also be statistically subtracted.

%%%%%%%%%%%%%%%%%%%%%%%%%%%%%%%%%%%%%%%%%%%%%%%%%%%
\section{Supernova $\nu_e$ detection and constraints}
\label{sec:SN nue detection}
%%%%%%%%%%%%%%%%%%%%%%%%%%%%%%%%%%%%%%%%%%%%%%%%%%%

We first discuss the typically-assumed range of supernova neutrino spectrum parameters and show spectra for some representative neutrino mixing scenarios.  We then calculate fits for the neutrino spectrum parameters and show the results for these and other cases.

\subsection{Calculated Detection Spectra}

Several cases can be considered for the initial spectra and how they are changed by neutrino mixing.  Our focus is on testing the $\nu_e$ sector.  We first note the two extreme cases that we want to differentiate and then mention some other possibilities.  There are also cases intermediate between the extremes we note.  We do not try to identify these cases in terms of active-flavor neutrino mixing scenarios, given the large uncertainties in the problem, especially in the initial neutrino spectra.  Our focus on improving the measurements, and the interpretation in terms of supernova emission and neutrino mixing will come once there is a detection.

Case (A) has $\langle E_{\nu_e} \rangle \approx 12$ MeV and $\langle E_{\nu_x} \rangle \approx$ 15 -- 18 MeV, i.e., there is a hierarchy of average energies between the flavors initially and neutrino mixing {\it has not} interchanged them (other assumptions are as above).

Case (B) has $\langle E_{\nu_e} \rangle \approx$ 15 -- 18 MeV and one flavor of $\nu_x$ has $\langle E_{\nu_x} \rangle \approx 12$ MeV (the other flavors of $\nu_x$ have $\langle E_{\nu_x} \rangle \approx$ 15 -- 18 MeV), i.e., there is a hierarchy of average energies between the flavors initially and neutrino mixing {\it has} interchanged them.

If the average energy of $\nu_x$ were large and mixing was effective at exchanging the spectra of antineutrinos instead of neutrinos, this would be evident in the $\bar{\nu}_e + p$ spectrum; this is disfavored by the SN 1987A data.  If all flavors had a low average energy, this would be evident in the $\bar{\nu}_e + p$ and $\nu + p$ spectra (because the $\nu + p$ channel is a neutral-current interaction, its yield is not changed by active-flavor mixing).  The yields of these and other channels can decide everything except the differences between Cases (A) and (B).  That's the open problem: What is the $\nu_e$ spectrum?

When the $\nu_e$ average energy is high, $\nu_e + ^{16}$O is a good detection channel; otherwise, it gives no useful signal because the yields are too small to be detected in the presence of backgrounds.  Typical average energies from supernova simulations are markedly lower than the values assumed a decade or two ago, so $\nu_e + ^{16}$O is now a much less favorable channel.  Besides $\nu_e + e^-$ in Super-K, there is no other detection channel in any existing detector that produces enough identifiable $\nu_e$ events when the average energy is low.  The yield of $\nu_e + e^-$ barely changes with changes in the average energy.  Another important change from a decade or two ago is that much lower energies can be detected, which improves the spectrum shape tests.

The main background for these reactions is the inverse beta events.  Some of these numerous events can be removed using an angular cut, but they still pose a formidable background.  This is shown in the left panel in Fig.~\ref{fig:IB} for the same average energies as in Fig.~\ref{fig:nu - e event spectrum}.  There are $\sim$ 128 $\nu_e + e^-$ events in the  40$^\circ$ cone, but this is swamped by $\sim 827$ inverse beta events.  In the absence of neutron tagging, it will be difficult to extract the $\nu_e$ signal from this background.

However, adding Gd to Super-K has a dramatic effect.  Assuming that the efficiency of neutron detection in a Gd-loaded Super-K is $90\%$, the inverse beta background will decrease to 83 events.  This strongly improves the detection prospects of $\nu_e$.  The $\bar{\nu}_e$ spectrum will be well measured by cleanly-identified inverse beta decay events using neutron detection by Gd.  This will allow statistical subtraction of the backgrounds due to $\bar{\nu}_e + e^-$ and the remaining $\bar{\nu}_e + p$ events.  Liquid scintillator detectors will measure the spectrum of $\nu_x$ from the $\nu + p$ channel, which is most sensitive to the flavors with the highest average energies.  This will allow statistical subtraction of the backgrounds due to the ${\nu}_x + e^-$ channel.  These subtractions only lead to modest increases in the uncertainties of the spectrum shown in the left panel of Fig.~\ref{fig:IB}.

The $\nu_e + ^{16}$O channel is only useful if the $\nu_e$ average energy is large, as otherwise the yield is too small.  Even for a $\nu_e$ average energy of 18 MeV, the backgrounds are still important.  There are $\sim 200$ signal events in the whole detector.  Excluding a forward cone of 25$^\circ$, $\sim 190$ events remain.  (The different choice of angle for the forward cone is because now we focus on higher energies, for which the angular resolution is better.)  In a detector without Gd, these would be overwhelmed by the $\sim$ 7071 inverse beta events, but neutron tagging by Gd will dramatically reduce this background.  This situation is shown in the right panel in Fig.~\ref{fig:IB}.  Again assuming an efficiency of 90\% in neutron tagging in a Gd-loaded Super-K, only $\sim$ 707 of the inverse beta events will remain.  This enormous reduction in background will greatly help in isolating the $\nu_e + ^{16}$O signal.

Fig.~\ref{fig:different average energies} shows how the detection spectra for $\nu_e + e^-$ and $\nu_e + ^{16}$O change with different assumptions about the $\nu_e$ average energy.  The yield for $\nu_e + e^-$ elastic scattering depends only weakly on the average energy but that for $\nu_e + ^{16}$O reaction changes dramatically.  See also Table~\ref{tab:yields}.  Both channels also have characteristic spectrum changes as the average energy changes, as shown in Fig.~\ref{fig:different average energies}.

\subsection{Fits for Neutrino Spectrum Parameters}

The detection spectra in Fig.~\ref{fig:IB} show that adding Gd to Super-K will greatly reduce backgrounds for supernova $\nu_e$.  We quantify the improvement in the determination of the $\nu_e$ spectrum parameters, $\langle E_{\nu_e} \rangle$ and $E_{\nu_e}^{\rm tot}$, by constructing a $\chi ^2$ and performing fits.  We use
\begin{equation}
\chi^2 = \sum _{\rm i}
\left(\frac{\mathcal{O}_{\rm i}(\langle E_{\nu_e} \rangle^0, E_{\nu_e}^{\rm tot,0}) - 
T_{\rm i}(\langle E_{\nu_e}\rangle, E_{\nu_e}^{\rm tot})}{\sigma_{\rm i}} \right)^2\,,
\label{eq:chisquared definition}
\end{equation}
where $\mathcal{O}_{\rm i}(\langle E_{\nu_e} \rangle^0, E_{\nu_e}^{\rm tot,0})$ are the numbers of events in each bin assuming the fiducial values of the parameters, $T_{\rm i}(\langle E_{\nu_e}\rangle, E_{\nu_e}^{\rm tot})$ are the same allowing different values, and $\sigma_{\rm i}$ are the uncertainties on the fiducial numbers.

Because all spectra except $\nu_e$ will be well measured separately, here we only need to fit for the $\nu_e$ spectrum parameters.  That is, we fit spectra like those in Fig.~\ref{fig:different average energies} after the remaining backgrounds shown in Fig.~\ref{fig:IB} have been statistically subtracted.  In the $\chi^2$ calculation, the numbers of events in the numerator are only those of the signals; the backgrounds affect the results by increasing the uncertainties in the denominator, which depend on the numbers of signal plus background events.

Put another way, if we set up a $\chi^2$ for the data before the statistical subtractions (Fig.~\ref{fig:IB} instead Fig.~\ref{fig:different average energies}), then the contributions from flavors besides $\nu_e$ would cancel in the numerator but not the denominator.  More precisely, those cancelations would occur only on average if typical statistical fluctuations were included.

To determine the allowed regions of parameters when a supernova is detected, we calculate $\Delta \chi^2$ relative to various assumed best-fit cases (using $\Delta \chi^2 = 4.6$ for two degrees of freedom to obtain the $90\%$ C.L. regions).

Our results indicate the likely size and shape of the allowed regions once a supernova is detected.  We make some reasonable approximations.  The uncertainties on the initial spectra and the effects of neutrino mixing are large, and the uncertainties on the neutrino cross sections are moderate.  In addition, we are considering only the time-averaged emission, whereas the average energies may vary during the burst.  The widths of the bins were chosen to have approximately equal numbers of $\nu_e + e^-$ events in each bin (at least $\simeq 10$ events per bin).  The numbers of events are then large enough that the Poisson uncertainties can be treated as Gaussian.

In Case (A) from above, there is a hierarchy between the average energies of different flavors, but their spectra {\it are not} interchanged by mixing, so the average energy of $\nu_e$ is low.  We take $\langle E_{\nu_e}  \rangle^0$ = 12 MeV and $E_{\nu_e}^{\rm tot,0}= 5 \times 10^{52}$ erg as fiducial parameters for this case.

If these are the true parameters of the supernova, then the left panel of Fig.~\ref{fig:chisquared} shows the likely precision with which the parameters would be reconstructed from the measured data in Super-K without or with Gd.  In this case, the primary constraint comes from the $\nu_e + e^-$ channel.  The $\nu_e + ^{16}$O channel does not have enough events relative to the backgrounds, though large values of $\langle E_{\nu_e}  \rangle$ can be excluded by the non-observation of a significant number of events.  The presence of Gd reduces the size of the allowed region significantly.  With both channels together, the allowed region would be centered on $\langle E_{\nu_e}  \rangle = 12$ MeV and would range from roughly 9 to 14 MeV.  Thus, with Gd, it would be possible to say that $\langle E_{\nu_e}  \rangle$ is different from $\langle E_{\bar{\nu}_e}  \rangle$ (which could be 15 MeV with $\sim 1\%$ precision).  This would not be possible without Gd, so this is an important difference.

In Case (B) from above, there is a hierarchy between the average energies of different flavors, and their spectra {\it are} interchanged by mixing, so the average energy of $\nu_e$ is high.  We take $\langle E_{\nu_e}  \rangle^0$ = 18 MeV and $E_{\nu_e}^{\rm tot,0}= 5 \times 10^{52}$ erg as fiducial parameters for this case.

If these are the true parameters of the supernova, then the right panel of Fig.~\ref{fig:chisquared} shows the likely precision with which the parameters would be reconstructed from the measured data in Super-K without or with Gd.  In this case, both channels have enough events to define allowed regions.  The steep energy dependence of the $\nu_e + ^{16}$O cross section gives a precise measurement of the average energy, though the large backgrounds and uncertainties mean that the total energy is not well determined.  As before, the presence of Gd improves the precision, especially for the $\nu_e + e^-$ channel.  With both channels together, the allowed region would be very small.  It would be easy to distinguish Case (A) and Case (B); Gd would greatly improve the significance of this comparison.

The presence of Gd is even more important when the neutrino average energies are closer to each other.  This is seen in some simulations, e.g., Ref.~\cite{Mueller:2012ak}, where $\langle E_{\nu_e} \rangle \approx$ 11 MeV, and $\langle E_{\bar{\nu}_e} \rangle \approx \langle E_{\nu_x} \rangle \approx$ 15 MeV.  Due to the less pronounced hierarchy, it will be much harder to distinguish scenarios like Case (A) and Case (B).

Fig.~\ref{fig:jointchisquared} shows our results (joint constraints using both channels) for the allowed regions of the $\nu_e$ spectrum parameters. In this case, the presence of Gd does not completely separate the 90\% C.L. contours, but it comes very close.  Without Gd, the two allowed regions cannot be separated at all, which would significantly degrade the ability to test the physics.

Recent long-term simulations show that the average energy of the neutrinos can change during the $\sim$ 10 sec emission time~\cite{Huedepohl:2009wh,Fischer:2011cy,Nakazato:2012qf}.  The average energy of $\nu_e$ typically changes from $\sim$ 12 MeV to $\sim$ 6 MeV.  For a detector like Super-Kamiokande, it might be difficult to detect this change of average energy.  For a future detector like Hyper-Kamiokande, which will have better precision the spectral properties (see later), such a difference could distinguished.

%%%%%%%%%%%%%%%%%%%%%%%%%%%%%%%%%%%%%%%%%%%%%%%%%%%
\section{Conclusions}
\label{sec:conclusion}
%%%%%%%%%%%%%%%%%%%%%%%%%%%%%%%%%%%%%%%%%%%%%%%%%%%

When the next Galactic supernova occurs, it is essential that we have a collection of detectors that can measure all neutrino flavors well.  Without this, we will be unable to fully address many important questions.  What is the total energy emitted in neutrinos and how is it partitioned among flavors?  Are the average energies of the various flavors different?  What do these results say about neutrino mixing and tests of exotic physics?  What do the differences between $\bar{\nu}_e$ and $\nu_e$ emission tell us about the neutron-to-proton ratio of the collapsing core?

At present, the only detector with a relatively large yield of $\nu_e$ events is Super-K.  Even so, this is only $\sim 10^2$ events using the $\nu_e + e^-$ channel.  If the average energy of $\nu_e$ is large enough, then the $\nu_e + ^{16}$O channel will have a comparable number of events.  The problem is the background of $\sim 10^4$ events from the inverse beta channel, $\bar{\nu}_e + p$.  This background can be reduced for $\nu_e + e^-$ using an angular cut, but not enough.

We demonstrate in detail a new technique to reduce backgrounds for both the $\nu_e + e^-$ and $\nu_e + ^{16}$O channels.  If Super-K adds Gd to improve the detection of $\bar{\nu}_e + p$, then $\sim 90\%$ of these events will be individually identified through detection of the neutron radiative capture on Gd in close time and space coincidence with the positron.  This would dramatically reduce backgrounds for other channels.  The remaining backgrounds can be statistically subtracted using independent measurements.

We show that the $\nu_e$ spectrum parameters, $\langle E_{\nu_e} \rangle$ (average energy) and $E_{\nu_e}^{\rm tot}$ (total energy), can be measured to $\sim 20\%$ or better if Super-K adds Gd.  This is a significant improvement over the capabilities of Super-K without Gd.  (For comparison, the precision for $\nu_x$ in existing scintillator detectors is comparable, and the precision for $\bar{\nu}_e$ in Super-K will be $\sim 1\%$.)  Further, this improvement could be the difference between being able to answer essential questions or not.  Unless very large liquid argon or liquid scintillator detectors are built, then we have no other way to adequately measure the $\nu_e$ spectrum.

Future extremely large water Cherenkov detectors like Hyper-Kamiokande would have a dramatic impact on detecting supernova $\nu_e$ using this technique.  The $\sim 25$ times larger volume would reduce the uncertainty on the $\nu_e$ parameters by factor of $\sim \sqrt{25} = 5$.  This requires using Gd in Hyper-Kamiokande, the prospects of which are prominently considered~\cite{Kearns:2013lea}.  (This would also require a new very large liquid scintillator detector~\cite{Wurm:2011zn,Li:2013zyd} to for improved measurements of $\nu_x$ using the $\nu + p$ channel.)

This new method of determining supernova $\nu_e$ would help improve our understanding of supernovae and neutrinos in many ways.  It provides yet another motivation for Super-K to add Gd.  Given how infrequent Galactic supernovae are, it is essential that the opportunity to measure $\nu_e$ well not be missed.

%%%%%%%%%%%%%%%%%%%%%%%%%%%%%%%%%%%%%%%%%%%%%%%%%%%%%%%%

\section*{Acknowledgments} 
We thank Andrea Albert, Basudeb Dasgupta, Shunsaku Horiuchi, Kohta Murase, Kenny C.Y. Ng,  Sergio Palomares-Ruiz, and the anonymous referee for discussions and helpful suggestions.   RL and JFB were supported by NSF Grant PHY-1101216 awarded to JFB.

\bibliographystyle{kp}
\interlinepenalty=10000
\tolerance=100
\bibliography{Bibliography/references}

\end{document}